\documentclass[twocolumn,letterpaper,10pt]{article}
\usepackage{iasted}
\usepackage{times}
\usepackage[dvips]{graphicx}

\usepackage{makeidx}  
\usepackage{wrapfig}
\usepackage{verbatim}
\usepackage{pifont}
\usepackage{color}
\usepackage{xcolor}
\usepackage{url}
\usepackage{bibspacing}

\definecolor{darkgray}{RGB}{050,050,150} 

\begin{document}

\title{BPM, Agile, and Virtualization Combine to Create Effective Solutions}

\author{
Steve Kruba \\
Northrop Grumman \\
15010 Conference Center Drive \\
Chantilly, VA, USA \\
email: steve.kruba@ngc.com \\
\and
Steven Baynes \\
Northrop Grumman \\
15010 Conference Center Drive \\
Chantilly, VA, USA \\
email: steve.baynes@ngc.com \\
\and
Robert Hyer \\
Northrop Grumman \\
15010 Conference Center Drive \\
Chantilly, VA, USA \\
email: bob.hyer@ngc.com \\
}

\maketitle
\thispagestyle{empty}



\noindent
{\textbf{\textit{Abstract---}}}{\textbf{The rate of change in business and  government is accelerating. A number of techniques
  for addressing that change have emerged independently to provide for automated
  solutions in this environment. This paper will examine three of the most popular of
  these technologies---business process management, the agile software development
  movement, and infrastructure virtualization---to expose the commonalities in
  these approaches and how, when used together, their combined effect results in
  rapidly deployed, more successful solutions.}
}
\vspace{2ex}

\noindent 
{\textbf{\textit{Keywords---Agile; \textsc{bpm}; business process
  management; Rapid Solutions Development; Virtualization; Workflow}}

%
\section{Introduction}
\label{sec:secIntroduction}
Supporting change in today's dynamic environment requires a strategy and tools that can
adapt to unforeseen events.  Such tools have evolved in three key areas independently in
response to this pressure.


 {\bf Business Process Management} (\textsc{bpm}) is both a management discipline and a set
  of technologies aimed at automating organizations' key business processes. Agility is a
  key feature of the products that support this market.

 {\bf Agile Software Development} is an approach for creating
  custom software and is designed to overcome some of the short-comings of more
  traditional approaches such as the waterfall methodology. It achieves agility through an
  iterative development approach that focuses on producing working software as quickly as
  possible.

{\bf Infrastructure Virtualization} has
  expanded from server virtualization to storage, network, and desktop virtualization. The
  emphasis is on providing computing resources transparently to users and applications so
  that solutions can be stood up and modified quickly, and managed more easily and
  effectively.


The term \emph{agile\/} has become popular for describing an important feature of modern
information technology architectures.  Agile within the context of each of the three
technologies described in this article has a slightly different connotation, but the
underlying principle remains the same. We will examine these similarities as well as the
differences.

We will examine each of these approaches separately within their agile context and will
discuss how in combination they are becoming increasingly important to creating successful
solutions. Examples from our experiences with our Northrop Grumman
e.POWER\raisebox{4pt}{\scriptsize\textregistered}\footnote{e.POWER has been providing
  solutions for government and commercial customers for over fifteen years and is a
  registered trademark of the Northrop Grumman Corporation.} \textsc{bpm} product will be
used to illustrate some of these ideas.

%
\section{Solutions}
\label{sec:secSolutions}
When acquiring new software tools, organizations typically begin by examining the feature
set of various products to determine which one is ``best'' at satisfying a set of
requirements. We can lose sight of the fact that what we're really looking for is a
\emph{solution} to a problem---not the tool itself.

This might seem like either an obvious or a nonsensical statement, depending on how you
look at it. Hasn't that always been the case with software development? you might ask. But
the fact of the matter is that deploying systems has gotten more complicated in recent
years. Quality issues, security issues, and compatibility issues have been given increased
visibility as organizations have been ``burned'' by not appreciating their importance.

\begin{figure*}
 \begin{center}
 \vspace{0pt}
  \includegraphics[width=6.7in]{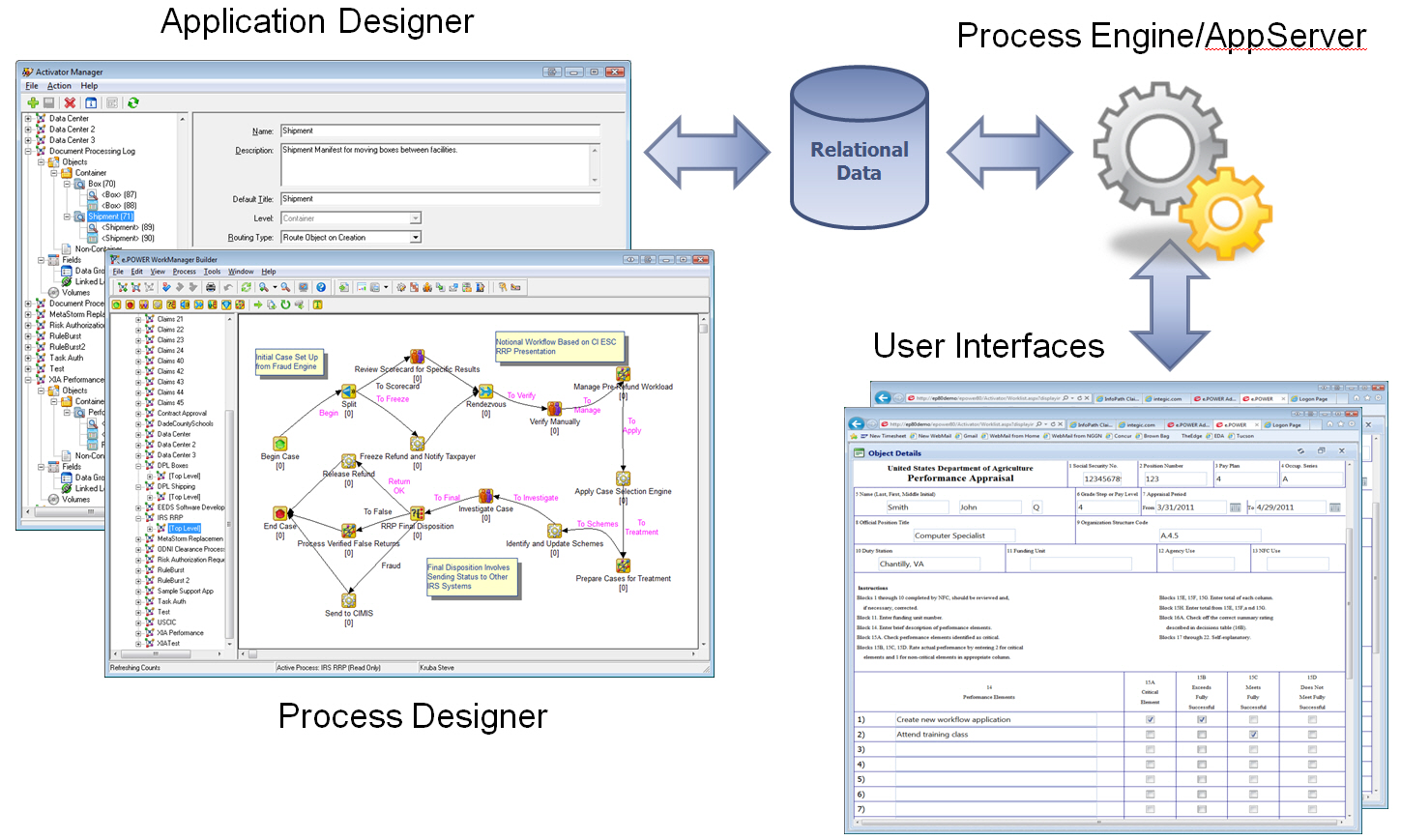}
  \caption{Model-Driven Architecture}
 \label{modeldriven}
 \vspace{-10pt}
 \end{center}
\end{figure*}

The effort involved in deploying finished solutions has become a significant part of the
solution creation process. Deployment costs can be comparable to development cost when
using model-driven tools in the \textsc{bpm} space. If those deployment (and support)
costs can be reduced through technologies such as virtualization, that is significant.

Combining an effective software development tool with a powerful methodology like the
agile process can be very beneficial. But \emph{writing\/} software is probably not the
best approach if there are solutions available that satisfy the requirements with
pre-written software.

And finally, when we step back to thing about ``solutions,'' we begin to focus on the
\emph{effectiveness\/} of the solutions produced. Key benefits for \textsc{bpm} and the
agile methodology are on how well the solutions produced meet the \emph{actual} needs of
their stakeholders. Very often with complex systems requirements, asking stakeholders to
define what they need is problematic because they simply do not have the experience to
articulate the details. Both \textsc{bpm} and agile are specifically designed to reduce
the risk of producing well-constructed solutions that are not effective at satisfying the
true requirements; i.e., producing a \emph{good} solution that is not the \emph{right}
solution.

It's worth noting that a significant percentage of business solutions today involve some
level of business process automation. Unlike other middleware components, rather than
being just another tool used in constructing the solution, \textsc{bpm} software
orchestrates the entire solution, even though it typically has to interact with many other
infrastructure components (e.g., other applications and services) that satisfy important
solution requirements.

In the next sections, we'll examine the three technologies in more detail. We will repeat
the key theme of how they improve the agility of the overall solution in the generic sense
(as opposed to the ``agile software'' sense) and hopefully gain insights into how to view
these engagements from an overall solution perspective.

%
\section{Business Process Management}
\label{sec:secBPM}

Business process management, or \textsc{bpm}, is a management discipline typically
supported by technology.\cite{gartnerbpm} The purpose of \textsc{bpm} is process
improvement. Software tools provide the technology base under which these goals are
achieved. A typical \textsc{bpm} solution is composed of tasks performed by people and
tasks performed by automated agents.

The \textsc{bpm} market, represented by over 100 vendors, is one of the fastest growing
software segments per Gartner Dataquest while business process improvement has been ranked
number 1 for the past five years by \textsc{cio}'s in the annual Gartner \textsc{cio}
survey.\cite{cio}

\textsc{bpms}'s such as e.POWER provide design environments that partition the work so
that users with diverse skill-sets can work independently when developing a
solution. Business users and business analysts play a major role in defining the business
process and associated rules and can use graphical interfaces for defining these
components. Graphics artists, rather than developers, can be used to design and implement
the layout of user interfaces. Software developers create customizations that access
legacy data from related applications, enhance the user interface by extending the
out-of-the-box functionality when necessary, and extend the functionality of the process
engine through exposed interfaces.

\begin{figure*}
 \begin{center}
 \vspace{0pt}
  \includegraphics[width=3.9in]{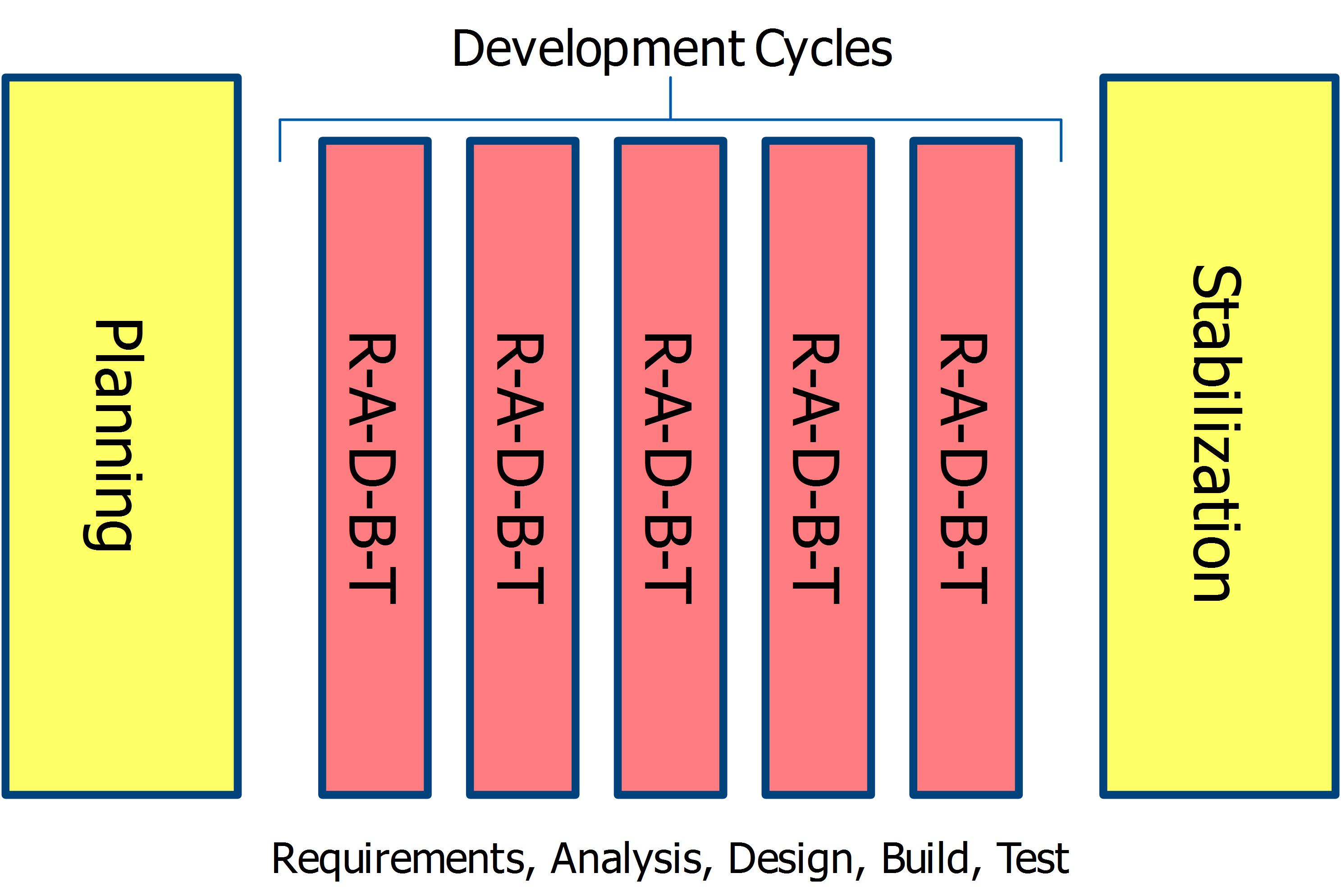}
  \caption{Our Agile/Iterative Approach}
 \label{cycles}
 \vspace{-10pt}
 \end{center}
\end{figure*}

A key differentiator of \textsc{bpms}'s is that they are \emph{model-driven}. Other
toolsets servicing the \textsc{bpm} market and other business software segments are either
parameterized, configuration-driven, or require writing custom software for the majority
of the functionality. The difference is that \textsc{bpms}'s provide this functionality
out-of-the-box.\cite{rsdbpm}. Parameterized or configuration-driven products are similar,
but the connection between the production instantiation of the model is not as direct as
model-driven products and only offer options that were pre-conceived by the product
developers. Model-driven products offer much greater flexibility.\footnote{Gartner has
  written a lot on this topic.\cite{gartnerchangeconcept} \cite{gartnerthree}
  \cite{gartnercomp} }

A pictorial representation of model-driven \textsc{bpm} is shown in Figure
\ref{modeldriven}. A graphical tool is used to define the business process, the results of
which are stored in a backend repository---often a relational database. Likewise an application 
designer is used to define an application that is ``process aware.'' This information
is used by the process engine for enforcing the business rules and routing rules
and by applications servers that drive the user interfaces. This same information is also available to
end-users as they interact with the system for managing and performing work.

Note that \emph{some} model-driven tools are used to define, not just the business
process, but also \emph{applications} that are process-enabled---the right-hand-side of
Figure \ref{modeldriven}. The user-interfaces needed to actually process work within the
business process are an important part of the solution, and being able to generate those
interfaces through configuration rather than coding is a very powerful
capability.\cite{rsdbpm}

\begin{center}
\colorbox{darkgray}{
  \parbox{2.9in}{\color{white}\textbf{The combination of capabilities
    provided by BPMS's fundamentally changes the way solutions are
    constructed in this problem space.}}}
\end{center}

A \textsc{bpms} product is purpose-built to create \textsc{bpm} \emph{solutions}. Within
the \textsc{bpms} framework, the features that are common to all process problems are
built into the product so that architects using the products simply deploy these pre-built
components, augmented by customized components needed to represent the uniqueness of each
particular solution. Frameworks such as service component architectures (\textsc{sca})
within a service oriented architecture (\textsc{soa}) are synergistic with \textsc{bpms}'s
for the customization components. \textsc{bpms}'s could be viewed as pre-compiled
frameworks.

This solutions-orientation is designed for rapid deployment and increased
effectiveness. Being able to construct these solutions quickly while using these
expressive tools to create more effective solutions is a key benefit. Effectiveness
is achieved by using the tools to improve requirements validation, design, and solution
creation, while improving quality and reducing risk.\cite{rsdbpm}

As we will see in the next sections, this agility can be amplified by other components of
the solution infrastructure.

\begin{figure*}
 \begin{center}
 \vspace{0pt}
  \includegraphics[width=5.7in]{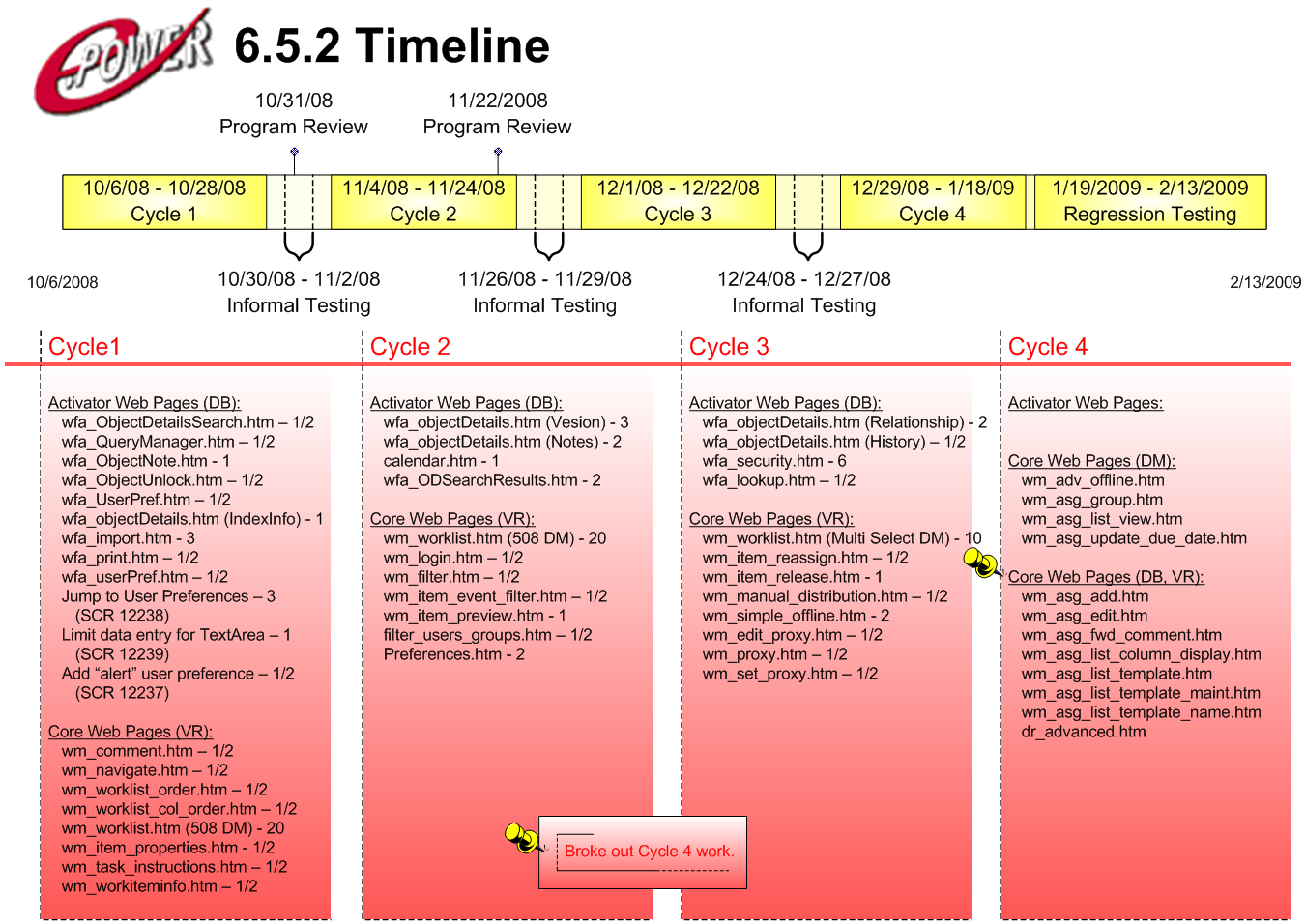}
  \caption{Release Timeline}
 \label{timeline}
 \vspace{-10pt}
 \end{center}
\end{figure*}

%
\section{Agile Software Development}
\label{sec:secAgile}

This section is not intended to be a how-to guide on agile software development---there
are a number of other excellent resources. We will, however, include a discussion of agile
principles to see how they relate to agility the other two technologies, illustrated by
our experience in developing our own product.

Agile software development is first and foremost an iterative methodology for producing
quality software. Agile development is characterized by frequent engagement of all
stakeholders, including customers, developers, quality personnel, and management. Agile
software development employs multiple development cycles to produce robust, testable
feature sets, followed by integrated system testing.

We especially like the following quotation on the goal of agile software
development.\cite{malotaux} This emphasizes the fact that all projects are time-bound and
helps to avoid the problem of scope-creep.

\begin{center}
\colorbox{darkgray}{
  \parbox{2.9in}{ \color{white}\bfseries At the end of a project we would
    rather have 80\% of the (most im\-por\-tant) fea\-tures 100\% done, than 100\% of
    all features 80\% done.}}
\end{center}


\subsection{e.POWER Agile Development}
\label{sec:secepagile}

To provide insight into agile development, we thought we would describe our first-hand
experience using agile development for product releases of the e.POWER product over the
past six years.

Our approach consists of three phases for delivering a quality product: planning, multiple
development cycles, and stabilization. Figure \ref{cycles} illustrates this approach.

The planning phase begins with defining a vision statement and is followed by developing a
list of features. Preliminary requirements are then collected and analyzed after which we
can focus on the most important requirements first. Consistent with the agile 
manifesto\cite{AgileManifesto}, planning requires frequent involvement with our stakeholders. We
kick off our planning sessions with a meeting with our advisory forum membership to be
certain that we collect their high-level input as well as their detailed requirements.

After completing the planning phase, we are in a position to commit to what we will do in
the iterative cycles. Each cycle is a mini-waterfall model, but much shorter, where we
finalize requirements, perform analysis, design the software, build it, and test it. Every
iteration or cycle contains a slice of the product, delivering small pieces of complete,
\emph{working} functionality.

The cycles provide opportunities for stakeholders that are not already part of the
development cycles to review completed functionality. Since each cycle produces
working software, demonstrations of that functionality are quite natural and simple to
produce. These reviews also provide the opportunity to reprioritize the features and
requirements list between cycles, since everyone is now more engaged and aware of the
evolving solution.

Figure \ref{timeline} provides an example of a release timeline of a past release of the
e.POWER product. We have used this template for several years to manage the process. This
one page summary of each release has been very effective at providing management,
developers, and testers with visibility into the process and managing to the schedule.

After completion of the final cycle, we enter the last, or stabilization phase. At this
point we perform complete system regression testing. Since we support multiple platforms
for each release, we do platform testing during this phase. The configuration control
board reviews the final requirements against our solution and documentation can now be
finalized. Our quality manager is responsible for leading our final ``total product
readiness''\footnote{Software products are comprised of much more than just
  \emph{software}. Total product readiness is a term that we use to include the full
  breadth of capabilities that must be delivered for a product release, including
  marketing collaterals, release announcement materials, installation scripts, on-line
  help, documentation, training materials, etc.}  process, which authorizes the product
for commercial release.

\subsection{Relationship of Agile to BPM Solution Creation}
\label{sec:secbpmagile}

Our experiences with agile development of our software product may be interesting, but how
does that relate to customers creating \textsc{bpm} solutions? They are related in two
ways:

\begin{dingautolist}{202}

\item \textsc{bpm} solutions typically involve writing \emph{some} custom software. To the
  extent this is minimal, the more robust and effective the solution can be.\cite{rsdbpm}
  But when significant customization is required, an iterative agile approach can be
  beneficial.

\item Perhaps more importantly, the methodology used in writing software for agile
  software development is very similar to the iterative approach that we have used over
  the years in creating e.POWER \emph{solutions}, including the model-driven aspects of
  the solution. The difference relates to code creation vs. solution creation. For agile
  \textsc{bpm} we are able to reduce the need to write custom software, replacing it with
  model manipulation---a non-programming effort.

\end{dingautolist}


%
\section{Virtualization}
\label{sec:secVirtualization}
Virtualization is a much over-hyped technology, but not without some justification. Data
centers world-wide are moving to virtualization to simplify operations, reduce hardware
costs, reduce cooling and energy costs, and expedite solution deployments.

Although virtualization gained recognition initially with data center servers and has been
in common usage for many years, virtualization has experienced much increased popularity
recently in the areas of storage, networking, and desktops.

The following sections provide a high-level summary of the important subtopics on
virtualization so that we can relate virtualization to our overall theme.

\begin{figure*}
 \begin{center}
 \vspace{0pt}
  \includegraphics[width=6.0in]{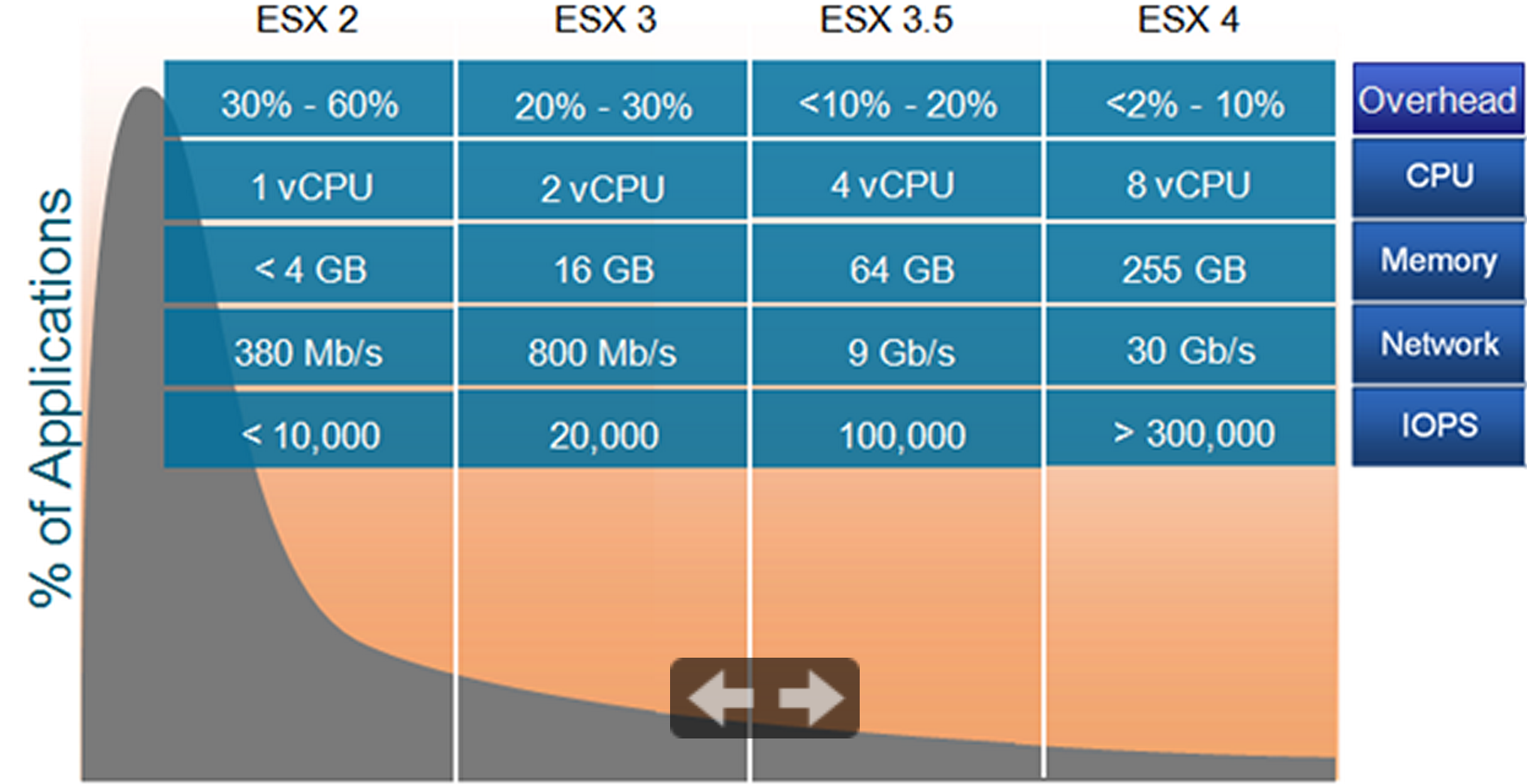}
  \caption{VMWare ESX Performance Improvements}
 \label{ESX}
 \vspace{-10pt}
 \end{center}
\end{figure*}

\subsection{Hardware}
\label{sec:secHV}

Hardware virtualization abstracts the physical computer hardware from the operating
system, allowing applications not originally designed for that combination to run on the
new, virtualized platform.

Improved management features greatly reduce the manpower needed to configure, secure,
backup, and operate virtualized servers than their physical counterparts. Furthermore,
server virtualization solutions form the basis for cloud computing, which can be viewed as
virtualization on steroids. Private clouds are virtualization platforms with even richer
management features. The importance to our discussion is that adding new solutions to
those environments becomes even easier yet, reducing operating costs.

Key aspects of virtual servers include higher availability (virtualization hardware can
bring up offline copies of the server automatically), faster provisioning of new servers,
automatic provisioning of new servers based on templates and security access rights, and
much simpler hardware upgrades since the virtual machine is independent of the
hardware. These are all aspects of agility that are important to our topic.

\subsection{Storage}
\label{sec:secSV}

In a manner similar to hardware virtualization, storage virtualization abstracts the
physical computer storage from the logical storage referenced in applications through the
operating system. The original impetus for storage virtualization may have been hardware
independence---the desire to be able to swap out one vendor's disk drives for another
vendor's when the old technology became obsolete. In general, features such as vendor
independence, over-provisioning, replication, pooling, improved utilization, snapshots,
etc., so significantly reduce operating costs that they typically offset any concern for a slight
reduction in performance.

\subsection{Network}
\label{sec:secNV}

Network virtualization uses software so that reconfiguring the physical network is not
necessary to implement operational changes. The major benefit of network virtualization is
simplified management of the network infrastructure. As new solutions are deployed, as new
hardware is fielded, and as work-patterns evolve to support changing business
requirements, network administrators can modify network configurations more easily than in
the past.

\subsection{Desktop}
\label{sec:secDV}

Desktop or client virtualization breaks the connection between users and their physical
desktop machines. Multiple users share instances of servers for their desktop computing
needs. Users require physical devices such as keyboards and monitors to interact with
their virtual desktop servers, but these devices can be components of a system running a
different operating system, or a purpose-built device limited to user interaction. In
either case, an individual user can be granted access to one or more virtual desktops for
specific work-tasks.

For some business cases, desktop virtualization offers significant benefits, but is not
necessarily optimal for all use cases. When appropriate, desktop virtualization offers
simpler and faster provisioning of new desktops, simplified desktop management in areas
such as backups and patch management, better security, and improved reliability,

\subsection{Performance Issues}
\label{sec:secPerf}

A discussion of virtualization would not be complete without consideration of the
performance implications. Virtualization provides an extra layer between applications and
the physical hardware resources that has an associated cost. Our discussions will center
around server virtualization.

For server virtualization, the hypervisor or Virtual Machine Monitor allows multiple 
operating systems to run concurrently on the machine hardware.
 }


{\bf Type 1} or bare-metal hypervisors run directly on the host hardware while
  the virtual operating system(s) run on top of them. These tend to be more efficient than
  Type 2 hypervisors. Type 1 hypervisors include Microsoft Hyper-V, VMWare ESX and ESXi,
  and Citrix Xen Server.

{\bf Type 2} hypervisors run on top of a host operating system such as Microsoft
  Windows, adding an additional level between applications and the hardware. Type 2
  hypervisors include VMWare workstation, VMWare server, and Microsoft Virtual Server.



The Virtual Insanity website has a useful graphic that illustrates the improvement in
server virtualization performance in the VMWare ESX product---a Type 1 bare-metal
hypervisor. As you can see from Figure \ref{ESX}, dramatic improvements have been made,
reducing overhead from 30--60\% in ESX 2 to 2--10\% in ESX 4.\cite{vinsanity} Note
corresponding improvements in network throughput and disk input/output (\textsc{iops}) as well. The
point is that for mission critical systems, you need to benchmark your virtualized
applications to insure adequate responsiveness and choose carefully to meet your needs.

Some Type 2 hypervisors are free and can be useful for some of your needs. For development
and support of the e.POWER product, our primary need is to provide support for many old
releases of the product, but since the activity levels are low, performance is not a
critical issue. For production customers, we provide target memory and \textsc{cpu}'s required
to support designated workloads, and those targets are somewhat ``diluted'' when deployed in
virtual environments. Depending on those activity levels, additional hardware may be
needed, the cost of which may well be recouped in reduced operating and support costs,
especially with the minimal overhead of the latest releases of Type 1 hypervisors.

On virtualized hardware, a key consideration is whether the storage is housed on internal
drives or external storage. On platforms such as VMWare, complete filesystems are
encapsulated when stored locally and are significantly less responsive than external
storage such as \textsc{san} storage or \textsc{nfs}. Server virtualization platforms have special drivers
for external storage that overcome this limitation.

\subsection{Our Experience}
\label{sec:secExp}

We thought it would be valuable to include information on our experience with
virtualization, primarily with hardware virtualization. The following chart summarizes 
the benefits we have seen from this transition over the past several years.

\begin{dingautolist}{202}

\item Ability to setup virtual machines quickly and move them or turn them on or off as
  needed

\item Reduced hardware investment and on-going power costs

\item More efficient use of hardware resources

\item Reduced labor expenses in moving virtual machines to new hardware---no operating
  system reinstallation necessary

\item Features such as ``snapshot'' facilitates testing of installation scripts and
  configuration

\item Entire virtual machines are backed up as a single file

\item Virtual machines are indistinguishable from physical machines from an end-user
  perspective

\end{dingautolist}


%
\section{Common Themes}
\label{sec:secCommon}
So what are some of the common themes we see in the three technologies presented above? At
a high level, they center around the concept of agility: providing the ability to create
solutions that are both quick to produce and adaptable to needs that evolve over
time. Speed is an important component but equally important is \emph{effectiveness},
emphasizing the overall development process from needs assessment through deployment. The
three approaches highlighted in this article are not the only ones that apply these
principles, but are three of the most visible today and touch on all aspects of solutions
in the business context. An organization that emphasized at least these three would be
well served.

Agility is about embracing change, knowing that user requirements will evolve as the
emerging solution provides greater visibility into the final product. \textsc{bpms} and
agile toolsets make it possible to iterate towards a solution because of the flexibility
they introduce into the creation process. We are often at a loss to express what the
ultimate solution needs to look like, but we can more readily recognize it when we see it.


A high level summary of principles that are shared among these three approaches are as
follows.  

\begin{dingautolist}{202}

\item People are the key to solutions. Technology has reached a level of refinement where
  we no longer should be optimizing bits and bytes, but optimizing people, including
  process participants, architects and developers, and support staff.

\item Engage stakeholders continuously throughout the solution development
  process. Continuous feedback with iterative evolution of the solution fundamentally
  improves the creation process.

\item Working software is the best visualization tool for working software. Modern
  software has become increasingly expressive, for which there is no substitute.

\item Task the right people for each aspect of solution creation based on domain
  knowledge. The tools now allow business people to participate in this process, allowing
  information technology professionals to focus on the \textsc{it}-aspects of solutions.

\item Modularity is a theme seen in all three of these technologies, allowing participants
  to easily conceptualize the current component of interest.

\item Virtualization makes infrastructure less of an impediment to productivity. People
  can more easily gain access to infrastructure resources in creating and managing
  solutions.

\end{dingautolist}


\section{Conclusions}
\label{sec:secConclusions}

Creating successful automated solutions is challenging in today's highly competitive
environment. Solutions must be conceived and implemented quickly in a manner that allows
them to adapt as needs change. For a large class of business problems, this requires the
capabilities of a business process management suite. \textsc{bpms}'s differentiate
themselves by their rapid solutions creation capabilities achieved through a model-driven
architecture. Since significant \textsc{bpm} projects require some custom software for
critical parts of the solution, agile principles are very compatible with \textsc{bpms}'s
and Northrop Grumman's e.POWER product software is developed using an agile software
development methodology.

Creating or evolving a solution rapidly is of little consequence if it cannot be fielded
in a like manner and this is where virtualization becomes important. The underlying
underpinnings of agility in each of these aspects of solution creation work together to
insure that solutions are effective and deployed in a timeframe that meets the needs of
their business customers.


\setlength{\bibspacing}{\baselineskip}
\bibliographystyle{plain}
\bibliography{AgileBPM}
%



%
\section*{Author Profiles}
\label{sec:secProfiles}
\noindent \textbf{Steve Kruba} is chief technologist for Northrop Grumman’s process-oriented 
commercial software products, including e.POWER, and a Northrop Grumman 
Technical Fellow. Steve has 42 years of experience developing software and
 solutions for customers. He holds a Bachelor of Arts in Mathematics and a 
Master of Science in Management Sciences from the Johns Hopkins University.

\noindent \textbf{Steve Baynes} is the department manager for the e.POWER product 
development team with extensive agile experience. He is a certified
 ScrumMaster, member of the Agile Alliance organization and speaks 
often on Agile development. As the manager for the e.POWER product,
 Steve works with business development, project implementation teams,
 and customers to continually improve the e.POWER product from both 
a feature and quality perspective.

\noindent \textbf{Bob Hyer} is chief architect for the e.POWER product development
team. Bob has over 30 years of experience developing software solutions
and software products for government and commercial customers.  He has
a Bachelors of Science in Business Management from Virginia Tech and a
Master of Science in Technology Management from American University.

\end{document}